\input amstex.tex
\input amssym.def
\input amssym.tex

\TagsOnRight
\hsize=6in
\vsize=8,5in
\parskip=4mm
\baselineskip=7mm

\tolerance=1000
\looseness=1
\parindent=8mm
\nobreak
\nopagenumbers

\font\rm=cmr10

\font\bf=cmbx10 
\font\sl=cmsl10

\magnification=1200

\centerline{\bf 	YANG-LEE EDGE SINGULARITY}
\centerline{\bf ON A CLASS OF TREELIKE LATTICES}

\vskip 1cm
\par
{
\centerline{\bf Milan
Kne\v zevi\'c and Sun\v cica Elezovi\'c--Had\v zi\'c}
\noindent {\sl Faculty of Physics, University of
Belgrade, P.O.Box 368, 11001 Belgrade, Serbia, Yugoslavia}

\par
}
\vskip 0.5cm  

{\bf Abstract.} The density of zeros of the partition function of 
the Ising model on a class of treelike lattices is studied. An exact
closed-form expression for the pertinent critical exponents is derived
by using a couple of recursion relations which have a singular behavior 
near the Yang-Lee edge.

\vskip 2.5cm
{\bf P.A.C.S numbers:} 05-50.+q, 64-60.Ak, 64-60.Fr, 75-40.Dy

\vfill
\eject

\footline={\hss \rm \folio\hss}
\pageno=2

It is well known that Ising model on finitely ramified fractal lattices
[1-2] can not display a phase transition at any finite temperature. 
This means that the free energy density $f(T,H)$ of such a model provides
an analytic function of temperature $T$ ($T>0$) and of a real magnetic
field $H$. Yang and Lee [3-4] were the first who pointed out that the free
energy, even of a finite Ising system, can exhibit a singular behavior 
if one allows the field $H$ to become complex $H=H'+iH''$. They showed that
all of the zeros of the partition function of a nearest-neighbor
ferromagnetic Ising model lie on the unit circle in the complex activity
plane $y=\exp(-2H/k_{_B}T)\equiv\exp(-2h)$. In the thermodynamic limit
these zeros are expected to condense, yielding to a limiting density of zeros
$g(h'',T)$. For any fixed temperature there exists a pair $\pm iH_0(T)$ of
zeros lying closest to the real $H$ axis, commonly referred to as the
Yang-Lee (YL) edge singularities. These limiting nonanalytic points exert
the most direct influence on the behavior of $f(T,H)$ for real $T$ and
$H$, and it is very important, therefore, to understand the nature of the
associated singularity in the density of zeros near the YL edge.

The calculation of the limiting density of zeros is generally a highly
nontrivial problem and little is known about its behavior near the YL
edge, let alone about its actual form on the whole region of interest.
It is widely accepted, however, that the density of zeros near the edge
exhibits a power-law behavior: 
$g\sim \vert h''-h_0(T)\vert^\sigma,\quad\text{as}\quad  h''\to
h_0(T)$, 
where $\sigma$ is the YL edge singularity critical exponent. 
Since in this case exists only a single relevant variable [5-8],
all other YL critical exponents can be expressed in terms of $\sigma$. In
particular, the correlation length critical exponent $\nu_c$, which
describes the spatial decay of the two-spin correlation function near the 
edge, $\xi_{YL}\sim \vert h''-h_0(T)\vert^{-\nu_c}$, 
can be related to $\sigma$ ($\sigma=d\nu_c -1$, with $d$ being the
dimensionality of underlying lattice).  
It has been suggested [6-8] that the value of this exponent depends only
on the dimensionality of the lattice and is independent of temperature for
all $T$ above the critical temperature. In $d=1$, the problem can be
solved exactly which yields to 
$\sigma=-1/2$ for $T>0$. In spite of the fact that $g(h'',T)$ is not known
exactly for the two-dimensional Ising ferromagnetic, it has been predicted  
$\sigma=-1/6$ in $d=2$ [9-10]. It is believed that $\sigma$ maintains its
mean-field value $\sigma=1/2$ above the upper critical dimension $d=6$
[7]. Only approximative analytical and numerical results are
available for the intermediate values of space dimensionalities [7-8].

The original YL circle theorem [3-4] is independent of the topological
structure of the lattice and should also apply to the appropriate models of
certain non-homogeneous systems (diluted ferromagnetics, for
example). In order to better understand certain aspects of their critical
behavior, few studies [11-12] of the density of zeros of the Ising
ferromagnetics on a variety of deterministic fractals have been
performed so far. It is shown that in this case the density of zeros
exhibits a scaling form near the YL edge which is more complicated than a
pure power law. Using the exact recursion relations nontrivial values of
the edge exponents $\sigma$ were found. Most of these values were estimated 
only numerically, due to the absence of a pertinent fixed point [12]
when one uses the standard decimation transformation approach.
To overcome this difficulty, one can try to construct a more liable  
renormalization-group scheme possessing the appropriate YL fixed point --
the block-spin transformation, for example. Unfortunately, such an
approach is usually much more elaborate, and, as soon as we know, only few
exact calculations are presented so far [11]. 
In this Letter we develop an approach which allows us to 
obtain the exact closed-form expressions for 
the edge exponents of Ising models on finitely
ramified fractal lattices. It turns out that relevant recursion relations
themselves have a singular behavior near the YL edge fixed point, which is
reminiscent of a singular structure of corresponding mappings appearing in
some recently studied Gaussian models [13-15]. 

Consider the nearest-neighbor ferromagnetic Ising model with a uniform
magnetic field on a family of treelike lattices shown in Fig. 1. Each
lattice of this family can be labeled by an integer $p$, which represents
the maximum coordination number of the lattice. This parameter can take
all values from 2 to $\infty$, with $p=2$ corresponding to a simple
one-dimensional chain. Let us note that each of these lattices have a finite
fractal dimension, $D=\ln p/\ln 2$, which make them very different from a
Cayley tree lattice.
The partition function of the Ising model of an $r$-th order lattice can
be written as a combination of only three partial partition functions:
$Z_1^{(r)}=Z^{(r)}(+,+)$, $Z_2^{(r)}=Z^{(r)}(-,-)$, and
$Z_3^{(r)}=Z^{(r)}(+,-)$, where $Z^{(r)}(+,+)$, 
for example, denotes the partial partition function with two outer Ising
spins [see Fig. (1b)] being fixed in the 'up state'. It is convenient to
express the resulting recursion relations in terms of two reduced variables 
$z_2=Z(+,-)/Z(+,+)$ and $z_3=Z(-,-)/Z(+,+)$. Thus, for a lattice of index
$p$, we have found
$$
\aligned
z_2'&=\frac{z_2^2(z_2+z_3)^{p-2}+y^{p-1}z_3^2(1+z_3)^{p-2}}
{z_3^2(z_2+z_3)^{p-2}+y^{p-1}(1+z_3)^{p-2}},\\
z_3'&=z_3\frac{z_2 (z_2+z_3)^{p-2}+y^{p-1}(1+z_3)^{p-2}}
{z_3^2(z_2+z_3)^{p-2}+y^{p-1}(1+z_3)^{p-2}},
\endaligned\tag 1
$$ 
with $y=\exp(-2h)$. To obtain an $r$-th order partition function
$z_i^{(r)}$ ($i=1,2$) one has to iterate the above recursion relations $r$
times, starting with the following initial conditions:  
$z_2^{(0)}=y^2$ and $z_3^{(0)}=xy$,  
where $x=\exp(-2K)$ (here $K=J/k_{_B}T>0$ stands for standard ferromagnetic
interaction strength).
 It is easy now to express various
derivatives of the free energy density in terms of these variables and
their derivatives with respect to a suitable variable. For example, the
average magnetization $M^{(r)}$ per spin is given by 
$$ 
{\Cal N^{(r)}}M^{(r)}=\frac{t_1+t_2+2t_3}{1+z_2+2z_3},\tag 2
$$ 
where we have omitted the iteration index $r$ on the right-hand side of
(6), and ${\Cal N^{(r)}}$ represents the
number of sites of the underlying lattice (${\Cal N^{(r)}}=p^r+1$, for a
lattice of index $p$). In the above formula $t_1$, $t_2$, and $t_3$
denote the scaled derivatives $t_i^{(r)}={(\partial Z_i^{(r)}/\partial
h)}/Z_1^{(r)}$, $i=1,2,3$, and they also can be calculated by a recursive
procedure. In a similar way, one can express the two-point correlation
function $G^{(r)}$ in terms of $z_2^{(r)}$ and $z_3^{(r)}$ 
$$
G^{(r)}=<S_1 S_2>-<S_1><S_2>=\frac{4(z_2-z_3^2)}{(1+z_2+2z_3)^2}.\tag 3
$$
The density of zeros $g(h'',T)$ is known [3] to be given by the
limiting behavior of the real part of $M/\pi$ as $h'\to 0$, so that we may
focus our attention here on the asymptotic behavior of the above variables
near the edge.

As it has been emphasized, for any $T>0$ there 
exists a strip $\vert H''\vert<H_0(T)$ inside of which the density of
zeros vanishes. The limiting value $H_0(T)$ can be determined numerically,
by a study of the magnetization (2) and the correlation function (3). Such
an analysis reveals that, for $H'=0$ and $H''\to H_0(T)$, $z_2$ and $z_3$
iterate toward the fixed point 
$$
z_2^*=(-y_0)^{\frac{2(p-1)}{p}},\quad z_3^*=-(-y_0)^{\frac{p-1}{p}},\tag 4
$$
where $y_0=\exp(-2ih_0)$ depends on $x$ (for $x=1/2$, we have found
$h_0=0.328\,648\,956\dots$ and $h_0= 0.152\,769\,589\dots$ for $p=3$ and $p=5$, respectively).
As it can be verified, however, recursion 
relations (1) become singular at this point, leading to a failing of the
common fixed-point analysis. This puzzle is quite similar to the one
encountered recently in the studies of critical properties of ideal
polymer chains on fractal space [13-15].  In particular, both
numerical and analytical examination of the above recursion relations
reveals the existence of an invariant 'line' $z_3=z_3(z_2)$, all points of
which are attracted by (4).  An asymptotic equation of this line, valuable
near the fixed point, can be expressed as an expansion over
the small variable $\delta z=z_3^2-z_2$ (note that at the fixed point one has
$z_2=z_3^2$). Thus we obtain
$$ 
z_3=z_3^*+c_1\sqrt{z_3^2-z_2}+c_2(z_3^2-z_2)+O[(z_3^2-z_2)^{3/2}],\tag 5
$$
where the coefficients $c_1$ and $c_2$ are given by
$$
\aligned
c_1&=\frac{1+\sqrt{1+4p}}{2p},\\
c_2&=\frac{p^2-2p-1+(3p^2-6p-1)z_3^*+[p^3-3p-1+(3p^3-2p^2-7p-1)z_3^*]c_1}
{2p^2z_3^*(1+z_3^*)(2+c_1+2pc_1)}.
\endaligned\tag 6
$$
Using the above formulas one can show that along the invariant line, near the
fixed point, $\delta z$ renormalizes according to the law: 
$\delta z'=2\delta z/(1+2p+\sqrt{1+4p})$. This enables us to describe the
way in which $z_3^{(r)}$ approaches its fixed point value: 
$$
z_3^{(r)}-z_3^*\sim \left(\frac{2}{1+\sqrt{1+4p}}\right)^{r},\quad r\gg
1.\tag 7
$$
In a similar way we can extract the leading asymptotic behavior of 
$\partial z_2^{(r)}/\partial h$ and $\partial z_3^{(r)}/\partial h$ at 
the edge. Indeed, it is easy to see that these derivatives satisfy a
couple of linear recursion relations, an analysis of which reveals the
following behavior 
$$
\left.\frac{\partial z_2^{(r)}}{\partial h}\right\vert_{h=ih_0}\sim
\left.\frac{\partial z_3^{(r)}}{\partial h}\right\vert_{h=ih_0}\sim 
2^r.\tag 8
$$

In the above established asymptotic relations we have supposed $h=ih_0$. 
In fact, they also hold for a finite but very small value of $\delta
h= h''-h_0$ ($0<\delta h\ll 1$), provided $r\lesssim r_0$,
where $r_0\gg 1$ is the number of iterations one can make along the
invariant line before going away from it (starting with the initial
conditions in which $y=y_0\exp(-2i\delta h)$). This number depends on
the value of $\delta h$ and can be estimated from the following obvious
relation: $z_3^{(r_0)}(\delta h)\approx z_3^* + \left.\frac{\partial
z_3^{(r_0)}}{\partial h}\right\vert_{\delta h=0}\delta h$. 
Thus, taking into account (7) and (8), we
find $\lambda^{r_0}\sim (\delta h)^{-1}$, where $\lambda=1+\sqrt{1+4p}$.
This enables us to express the YL correlation length as a function of $\delta h$:
$\xi_{YL}\sim 2^{r_0} \sim\exp(-\ln 2\ln (\delta h)/\ln\lambda)$, i.e.
$$
\xi_{YL}\sim {\delta h}^{-\nu_c}\quad\text{with}\quad \nu_c=\frac{\ln
2}{\ln\lambda}=\frac{\ln 2}{\ln(1+\sqrt{1+4p})},\tag 9
$$
independent of the temperature. 
We have also calculated the critical exponent $\nu_c$ for several values
of $p$, by using a numerical study of the correlation function (3),
and we have found an excellent agreement with (9).  
Let us note further that for $p=2$ the value of $\nu_c$ ($\nu_c=1/2$)
coincides with the known value for the Ising ferromagnetic chain, while
for $p=3$ our exact value of $\nu_c$ is very close to the one 
that has been obtained earlier by using somewhat different numerical
approach [12].  What is perhaps more interesting, our values of
$\nu_c$ for $p=2,3$ are the same as those found for the correlation length
critical exponent $\nu_{_G}$ of a simple zero-field Gaussian model on
the same lattices [13-15]. It is tempting, therefore, to see whether this
coincidence persists for all values of $p$. Our calculation [16], which follows
the lines of Ref. [13], shows that this is, really, the case ($\nu_c=\nu_{_G}$). 

In a similar way we have found: 
$ \left.t_1^{(r)}\right\vert_{h=ih_0}\sim 
\left.t_2^{(r)}\right\vert_{h=ih_0}\sim
\left.t_3^{(r)}\right\vert_{h=ih_0}\sim\lambda^r$.
Taking into account these results and (2), we can write $M^{(r_0)}\sim
t_1^{(r_0)}/{\Cal N^{(r_0)}}\sim (\lambda/p)^{r_0}$, which yields to
$$
M\sim g\sim \delta h^\sigma,\quad \sigma=\frac{\ln
p}{\ln(1+\sqrt{1+4p})}-1.\tag 10
$$ 
It is interesting to note that this expression can be rewritten in the
form of an expected scaling relation: 
$\sigma=D\nu_c-1$.  As it can be seen from the above formula,
$\sigma=\sigma(p)$ is an increasing function of $p$ and its limiting
value, $\sigma(p\to\infty)=1$, exceeds the above mentioned mean-field value
$\sigma=1/2$ for the homogeneous lattices. Note that $\sigma(p)$ changes
its sign at $p=6$, which means that singular part of the density of zeros
diverges near the edge if $p<6$, whereas it vanishes for all $p>6$. As 
can be noticed in Fig.2, where we plotted the density of zeros as a
function of $h''$ for several values of the lattice parameter $p$, the 
zeros are not homogeneously distributed on the unit circle but instead
have a striking Cantor-set structure with many gaps, especially for the low
values of $p$ (except, of course, for $p=2$). As $p$ increases,  
this fractal pattern become less and less pronounced, and it seems that,
for any small but finite $h'$, it acquires a rather regular limiting
form (see Fig. 2. for $p\gg 1$). For a fixed and finite temperature, the
value of the YL edge decreases with $p$ and one can show that $h_0\sim 1/p$ 
when $p\gg 1$, corresponding to the absence of any gap in the density of
zeros, in the limit when the lattice coordination number tends to infinity.
On the other hand, for a finite $p$ the density of zeros exhibits a scaling
form near the edge, similar to the one that has been discussed in details
in Ref. [11-12].  

In conclusion, we have presented  exact results for the density of
zeros in the complex activity plane for a family of treelike lattices. 
We have found that our recursion relations have a singular behavior 
near the edge. The technique we have used here to extract the leading
singularity can be, in principle, extended to all other finitely ramified 
fractal lattices and to other classical spin models.

\vfill
\eject
{\bf References}
\vskip3mm
\item{[1]} Y.~Gefen, A.~Aharony and B.~B.~Mandelbrot, J. Phys. A {\bf 16},
1267 (1983).
\item{[2]} Y.~Gefen, A.~Aharony, Y.~Shapir and B.~B.~Mandelbrot, J. Phys.
A {\bf 17}, 435 (1984). 
\item{[3]} C.~N.~Yang and T.~D.~Lee, Phys. Rev. {\bf 87}, 404 (1952).
\item{[4]} T.~D.~Lee and C.~N.~Yang, Phys. Rev. {\bf 87}, 410 (1952).
\item{[5]} D.~A.~Kurtze, M.S.c. Report No. 4184, Cornell University, 1979
(unpublished). 
\item{[6]} P.~J.~Kortman and R.~B.~Griffiths, Phys. Rev. Lett. {\bf 27},
1439 (1971).
\item{[7]} M.~E.~Fisher, Phys. Rev. Lett. {\bf 40}, 1610 (1978). 
\item{[8]} D.~A.~Kurtze and M.~E.~Fisher, Phys. Rev. B {\bf 20}, 2785
(1979). 
\item{[9]} D.~Dhar, Phys. Rev. Lett. {\bf 51}, 853 (1983).
\item{[10]} J.~L.~Cardy, Phys. Rev. Lett. {\bf 54}, 1354 (1985).
\item{[11]} M.~Kne\v zevi\' c and B.~W.~Southern, Phys. Rev. B {\bf 34},
4966 (1986).
\item{[12]} B.~W.~Southern and M.~Kne\v zevi\'c, Phys. Rev. B {\bf 35}, 5036
(1987). 
\item{[13]} M.~Kne\v zevi\' c and D.~Kne\v zevi\'c, Phys. Rev. E {\bf 53}, 
2130 (1996).
\item{[14]} A.~Maritan, Phys. Rev. Lett. {\bf 42}, 2845 (1989).
\item{[15]} A.~Giacometti, A.~Maritan, and H.~Nakanishi, J. Stat. Phys.
{\bf 75}, 669 (1994). 
\item{[16]} M.~Kne\v zevi\' c, (unpublished).

\vfill
\eject

{\bf Figure Captions}

Fig 1. (a) First two stages in the iterative construction of a $p=4$
treelike fractal lattice. (b) Schematic representation of an $r$-th stage
$p=3$ fractal lattice. In order to obtain a partition function
$Z_i^{(r)}(S_1,S_2)$, $i=1,2,3$, one has to fix the states
of any two outer Ising spins $S_1$ and $S_2$ (open circles) and perform
summation over the states of all remaining spins [including the central
spin (black circle)]

Fig 2. Density of zeros $g(h'',T)$ as a function of the imaginary field
$h''$ for several values of $p$. In all cases temperature corresponds to
$x=1/2$ and the real part of the field is $h'=0.005$.

\bye